\documentstyle[11pt,newpasp,twoside]{article}
\def\edcomment#1{\iffalse\marginpar{\raggedright\sl#1\/}\else\relax\fi}
\reversemarginpar

\begin{document}
\title{Seven problems related to the determination of the primordial 
helium abundance}
\author{Manuel Peimbert, Antonio Peimbert} \affil{Instituto de
Astronom\'{\i}a, Universidad Nacional Aut\'onoma de M\'exico; Apdo. postal
70--264; Ciudad Universitaria; M\'exico D.F. 04510; M\'exico.}

\author{Valentina Luridiana} \affil{Instituto de
Astrof\'{\i}sica de Andaluc\'{\i}a (CSIC), Camino Bajo de Hu\'etor 24,
18008 Granada Spain.}

\author{Mar\'{\i}a Teresa Ruiz} \affil{Departamento de
Astronom\'{\i}a, Universidad de Chile ; Casilla 36-D; Santiago; Chile.}

\begin{abstract}
Recent advances on the quest to determine the primordial or
pregalactic helium abundance, $Y_p$, are reviewed. There are seven
problems affecting the He/H abundance determinations of H~{\sc{ii}}
regions that are briefly discussed: the underlying absorption lines in
the observed spectra, the ionization, temperature, and density
structures of each object, the collisional contribution to the
intensity of the He~{\sc{i}} and H~{\sc{i}} lines, the optical
thickness of the He~{\sc{i}} lines, and the extrapolation to derive
$Y_p$ based on the $Y$ and O/H values.
\end{abstract}

\section{Overview}

The determination of helium abundances in H~{\sc{ii}} regions and
their extrapolation to derive $Y_p$ have become an important field of
research due to its relation to cosmology, to the chemical evolution
of galaxies, and to the study of the physical conditions inside
H~{\sc{ii}} regions.  Roberto and Elena Terlevich have done
significant work in two areas related to this field: the search for
metal poor H~{\sc{ii}} regions to derive their helium abundance, and
the gathering of high quality spectroscopic data to derive the $Y_p$
value (e.g. Terlevich et al. 1991a,b; Pagel et al. 1992).  Two recent
review papers on $Y_p$ are those by Steigman (2002) and Luridiana
(2002). This review will be mainly devoted to seven problems that
affect the $Y_p$ determination.

\section{Underlying Absorption Lines}

The observed spectra of giant extragalactic H~{\sc{ii}} regions is
produced by a combination of nebular emission and stellar
emission. The stellar emission includes a continuum with the H and He
lines in absorption.  If the underlying absorption is not taken into
account the intensity of the H and He emission lines will be
underestimated. The correction for underlying absorption is larger, and
consequently the associated errors, for objects with lower H$\beta$
equivalent width.

There are two ways to minimize the errors introduced by this problem,
a) to have enough angular resolution to be able to avoid the light of
the early type stars in the observing slit, this can be done only for
H~{\sc{ii}} regions of the local group (e.g. Peimbert, Peimbert, \&
Ruiz 2000; Peimbert 2003), or b) to have a good model of the starburst
and produce the expected stellar spectrum, like the work carried out
by Gonz\'alez-Delgado, Leitherer, \& Heckman (1999). Further
extensions of the work by Gonz\'alez-Delgado et al. are needed to
cover other metallicities and to include the equivalent widths of
other He~{\sc{i}} lines like $\lambda\lambda$ 5876 and 6678.

\section{Ionization Structure}

To determine very accurate He/H values of a given H~{\sc{ii}} region
we need to consider its ionization structure.  The total He/H value is
given by:
\begin{eqnarray}
\frac{N ({\rm He})}{N ({\rm H})} & = &
\frac {\int{N_e N({\rm He}^0) dV} + \int{N_e N({\rm He}^+) dV} + 
\int{N_e N({\rm He}^{++})dV}}
{\int{N_e N({\rm H}^0) dV} + \int{N_e N({\rm H}^+) dV}},
						\nonumber \\
& = & ICF({\rm He})
\frac {\int{N_e N({\rm He}^+) dV} + \int{N_e N({\rm He}^{++}) dV}}
{\int{N_e N({\rm H}^+) dV}}
\label{eICF}
.\end{eqnarray}

For objects of low degree of ionization it is necessary to consider
the presence of He$^0$ inside the H$^+$ zone, while for objects of
high degree of ionization it is necessary to consider the possible
presence of H$^0$ inside the He$^+$ zone. For objects of low degree of
ionization $ICF({\rm He})$ might be larger than 1.00, while for
objects of high degree of ionization $ICF({\rm He})$ might be smaller
than 1.00. The $ICF({\rm He})$ problem has been discussed by many
authors (e.g. Shields 1974; Stasi\'nska 1983; Pe\~na 1986;
V\'{\i}lchez \& Pagel 1988; Pagel et~al. 1992; Armour et~al. 1999;
Peimbert \& Peimbert 2000; Viegas, Gruenwald, \& Steigman 2000; Viegas
\& Gruenwald 2000; Ballantyne, Ferland, \& Martin 2000; Sauer \&
Jedamzik 2001; Gruenwald, Steigman, \& Viegas 2002).

The deviations from unity in the $ICF({\rm He})$ value occur in and
near the ionization boundary of a given H~{\sc{ii}} region, therefore
those H~{\sc{ii}} regions that are density bounded in all directions
have an $ICF({\rm He})$ = 1.00. Rela\~no, Peimbert, \& Beckman (2001)
from the spectral types of the ionizing stars of NGC~346 find that
about half of the ionizing photons escape the nebula favoring an
$ICF$(He) = 1.00, they support this result by fitting the observed
line intensities with a photoionization model. From the work by
Zurita, Rozas, \& Beckman (2000) on the ionization of the diffuse
interstellar medium in external galaxies it is expected that a large
fraction of the ionizing photons escapes from the most luminous
H~{\sc{ii}} regions, which favors the assumption that the $ICF$(He) is
very close to 1.00. Luridiana, Peimbert, \& Peimbert (2003), from
tailor made models of I~Zw~18, SBS~0335-052, and~Haro 29 also find
$ICF$(He) values very close to 1.00.

\section{Temperature Structure}

$T$(4363/5007) has been used often to determine the helium abundance,
under the assumption of constant temperature. However, from
photoionization models of H~{\sc{ii}} regions, it has been found that
the mean temperature variation, $t^2$, is in the 0.002 to 0.03 range,
with typical values around 0.005 (e.g. Gruenwald \& Viegas 1992;
Kingdon \& Ferland 1995; Pe\'rez 1997). Moreover from photoionized
models it is found that in very metal poor H~{\sc{ii}} regions the
zones where the [O~{\sc{iii}}] lines originate are several thousand
degrees hotter than the regions where the [O~{\sc{ii}}] lines
originate, while the He~{\sc{i}} lines originate in both regions
(e.g. Stasi\'nska 1990; Peimbert, Peimbert \& Luridiana 2002;
Luridiana et al. 2003).

{From} observations of galactic and extragalactic H~{\sc{ii}} regions
there is growing evidence that temperature variations are higher than
those predicted by chemically homogeneous photoionization models, for
example: a) the observed $T$(4363/5007) values are considerably higher
than those computed by photoionization models (e.g. Stasi\'nska \&
Schaerer 1999; Luridiana et al. 2003), b) the Balmer temperatures for
Magellanic Cloud H~{\sc{ii}} regions are considerably smaller than the
$T$(4363/5007) values (Peimbert et al. 2000; Peimbert 2003), c) under
the assumption of a constant temperature the C and O abundances
derived from the recombination lines of C~{\sc{ii}} and O~{\sc{ii}}
are considerably higher than those derived from the collisionally
excited lines of C~{\sc{iii}} and O~{\sc{iii}} (e.g. Peimbert et
al. 1993; Esteban et al. 1998, 2002; Peimbert 2003; Tsamis et
al. 2003), d) the self consistent method employed by Peimbert et
al. (2000) to derive the He$^+$/H$^+$ value also indicates that the
representative temperature for the He~{\sc{i}} lines,
$T$(He~{\sc{ii}}), is considerably smaller than the representative
temperature for the [O~{\sc{iii}}] lines, $T$(O~{\sc{iii}}).  All
these results imply that $T$(4363/5007) is an overestimate of
$T$(He~{\sc{ii}}). The self consistent method to derive He$^+$/H$^+$
requires a higher density for a lower temperature, and the higher the
density the higher the collisional contribution to the He~{\sc{i}}
line intensities, which results in a lower He$^+$/H$^+$ value.  This
problem has been amply discussed by Peimbert et al. (2002).

\section{Density Structure}

To produce a good photoionization model and to estimate the
collisional excitation of the He~{\sc{i}} lines a good density structure
is needed. H~{\sc{ii}} regions show very large density
fluctuations, this is apparent in any high resolution image of
those giant extragalactic H~{\sc{ii}} regions that have been used to
determine the pregalactic helium abundance. By comparing the
root mean square density with those densities derived from
forbidden lines it is possible to estimate the filling factor,
$\epsilon$, which is given by
\begin{equation}
N_e^2(rms)={\epsilon\; N_e^2(FL)}.
\end{equation}

Typical values of $\epsilon$ are in the 0.1 to 0.001 range
(e.g. Luridiana et al. 1999, 2003). There are five sets of forbidden
lines that have been used to estimate the density: [S~{\sc{ii}}],
[O~{\sc{ii}}], [Fe~{\sc{iii}}], [Cl~{\sc{iii}}], and [Ar~{\sc{iv}}].
Each set samples a different part of the H~{\sc{ii}} region, typically
the [S~{\sc{ii}}] density samples the outermost 2-4\% part,
[O~{\sc{ii}}] and [Fe~{\sc{iii}}] sample the 10 to 15\% outer parts,
[Cl~{\sc{iii}}] samples about 85\% of the object, and [Ar~{\sc{iv}}]
samples the innermost 2-4\%. Unfortunately for most of the well
observed H~{\sc{ii}} regions there are only [S~{\sc{ii}}] densities
available.

Often the [S~{\sc{ii}}] density has been used to determine the
collisional effect on the He~{\sc{i}} lines, this has to be considered
as a first approximation, but not good enough to derive very accurate
$Y$ values. For example for NGC~2363 the [S~{\sc{ii}}] density is
smaller than the [Ar~{\sc{iv}}] density (Per\'ez, Gonz\'alez-Delgado,
\& V\'{\i}lchez 2001), for a position in 30 Dor the [O~{\sc{ii}}] and
[Fe~{\sc{iii}}] densities are smaller than the [S~{\sc{ii}}] density
(Peimbert 2003), and for NGC~346 it has been found that a self
consistent method to derive the He abundance, based on 9 He~{\sc{i}},
lines requires a higher density than that provided by the
[S~{\sc{ii}}] lines (Peimbert et al. 2000).

The [S~{\sc{ii}}] lines, in addition to be representative of only a
small fraction of the H~{\sc{ii}} region, have the problem that are almost
insensitive to values of the density smaller than about 100 cm$^{-3}$.
Whenever possible we recommend the use of the [Fe~{\sc{iii}}] lines instead 
of the [S~{\sc{ii}}] lines on two grounds: they represent a larger fraction 
of the H~{\sc{ii}} region and they are very sensitive at low densities, 
specially the 4986/4658 ratio (Keenan et al. 2001).

\section{Collisional Excitation of the He~{\sc{i}} and H~{\sc{i}} Lines}

Recent expressions to correct for the collisional excitation of the
He~{\sc{i}} lines have been presented by Kingdon \& Ferland (1995) and by
Benjamin, Skillman, \& Smits (1999).

Davidson \& Kinman (1985) were the first to point out the relevance of
collisional excitation of the Balmer lines from the ground level of
the H atom.  Additional discussion and estimates of the relevance of
this process were presented by Skillman \& Kennicutt (1993),
Stasi\'nska \& Izotov (2001), and Peimbert et al. (2002). The
importance of this effect is proportional to H$^0$/H$^+$ and to the
Boltzmann factor for collisional excitation. In extremely metal poor
objects, that have high electron temperatures, the contribution of
this effect to the intensity of H$\beta$ can reach values of a few per
cent.  Since H$^0$/H$^+$ can not be derived directly from observations
we need tailor made models for each H~{\sc{ii}} region to properly
estimate the importance of this effect. For objects with $T_e > 17
000$ K probably the collisional excitation of the Balmer lines
introduces the highest source of error in the $Y$ determination
(Luridiana et al. 2003).

\section{Optical Thickness of the He~{\sc{i}} Triplet Lines}

The He~{\sc{i}} line intensities of the triplet system
are affected by the $2^3$S level optical depth. Therefore the
triplet line intensities have to be corrected for this effect to
derive accurate He/H abundance ratios. Benjamin, Skillman, \& Smits (2002)
have estimated this effect for the case of spherical geometry, they
conclude that their computations can be applied to observations for 
values of $\tau_{3889}$ smaller than 2. 

There are H~{\sc{ii}} regions with values of $\tau_{3889}$ larger than 2
and there are H~{\sc{ii}} regions that deviate considerably from spherical
symmetry, the triplet lines of these objects need to be corrected
for optical depth effects.

What is usually done to correct the line intensities is to determine
$\tau_{3889}$ from the ratio of the most affected triplet line to a
singlet line (the singlet lines are independent of this effect), and
from the derived $\tau_{3889}$ apply the spherically symmetric
solution to the $\lambda\lambda$ 5876 and 4471 lines to correct their
intensities, in general a small correction.

The four more sensitive lines to $\tau_{3889}$ are $\lambda\lambda$
3188, 3889, 4713, and 7065; two of them, $\lambda\lambda$ 4713 and
7065, increase in intensity with increasing $\tau_{3889}$, while two
of them, $\lambda\lambda$ 3188 and 3889, decrease.

For a region of 30 Doradus Peimbert (2003), based on the computations
of Benjamin et al. (2002), found two values of $\tau_{3889}$: a value
of 4.4 based on $\lambda\lambda$ 4713 and 7065 and a value of 10.5
based on $\lambda\lambda$ 3188, 3889. This result indicates that the
computations for spherical geometry by Benjamin et al. (2002) do not
apply to the observed region of 30 Dor.

Peimbert (2003) has suggested to use the $\tau_{3889}$ value derived
from the $\lambda\lambda$ 4713 and 7065 line intensities together with
the computations by Benjamin et al. (2002) to correct $\lambda\lambda$
5876 and 4471 for all objects. His suggestion is based on the
following argument: while the $\lambda \lambda$ 3188 and 3889 line
intensities depend on the optical depth along the line of sight, the
$\lambda\lambda$ 4713, 7065, 5876, and 4471 lines depend on atoms
absorbing the photons from all lines of sight and then re-emitting
them towards us, and thus depend on the average optical depth along
all angles. Therefore it is expected that the effective $\tau_{3889}$
will be the same for all lines whose flux intensity increases with
increasing $\tau_{3889}$; this effective $\tau_{3889}$ might be
different to that derived from $\lambda\lambda$ 3188 and 3889. For 30
Doradus the helium abundances derived from $\lambda\lambda$ 5876 and
4471 using the $\tau_{3889}$ obtained from $\lambda\lambda$ 4713 and
7065 are in excellent agreement with the abundances derived from 5
singlet lines which are not affected by this effect, supporting the
suggestion by Peimbert.

\section{$\Delta Y$/$\Delta O$}

To determine the $Y_p$ value from a given H~{\sc{ii}} region it is
necessary to estimate the fraction of helium present in the
interstellar medium produced by galactic chemical evolution. Often
$Y_p$ has been obtained from
\begin{equation}
\label{DeltaO}
Y_p  =  Y - O \frac{\Delta Y}{\Delta O},
\end{equation}
where all quantities are given by mass.  Some of the best $\Delta
Y$/$\Delta O$ values in the literature are presented in Table 1. The
second column includes irregular, H~{\sc{ii}}, and blue compact
galaxies. The observed value in the third column was derived by
comparing the abundances of M17 in the Galaxy (Peimbert,
Torres-Peimbert, \& Ruiz 1992; Esteban et al. 1999) with the
primordial helium abundance (Peimbert et al. 2002; Peimbert 2003).
The $\Delta Y$/$\Delta O$ observational values presented in Table 1
were derived under the assumption of temperature variations along the
line of sight; by assuming constant temperature the $\Delta Y$/$\Delta
O$ values become twice as large (Peimbert 2003).

\begin{table}
\begin{center}
\caption{$\Delta Y/\Delta O$ Values}
\begin{tabular}{lcc}
\\
\hline
\hline
\rule[-3mm]{0mm}{8mm}
Source & Irregulars & The Galaxy  \\ [0.2ex]
\hline
\\
Carigi et al. (1995), observations    & $4.5  \pm 1.0$    & ... \\
Izotov and Thuan (1998), observations & $2.7  \pm 1.2$    & ... \\
Peimbert (2003), observations         & $2.93 \pm 0.85$   & $3.57 \pm 0.67$  \\
Carigi et al. (1995), theory          &  2.95             & ... \\
Chiappini et al. (1997), theory       &  ...              & 3.15 \\
Carigi et al. (1999), theory          &  4.2              & ... \\
Carigi (2000), theory                 &  ...              & 2.9 - 4.6 \\
\\
\hline
\hline
\end{tabular}
\end{center}
\end{table}

The theoretical values for irregular galaxies presented in Table 1
were derived from closed box models and outflow models of well mixed
material. For O-rich outflows the models enter in contradiction with
the observed C/O values.  In Table 1 we present also the predictions
for the Galaxy based on infall models by Chiappini et al. (1997), and
Carigi (2000); the range in the $\Delta Y$/$\Delta O$ values derived
by Carigi comes from the use of seven different sets of stellar yields
present in the literature.

{From} Table 1 we conclude that, to derive $Y_p$, a $\Delta Y/\Delta
O$ = 3.5 $\pm 0.9$ is a good representative value to use in equation
(3).

\section{Other Problems}

The accuracy of the atomic parameters needed to derive the abundances
from the He~{\sc{i}} line intensities seems to be in the 0-3\%
range. A comparison of the He~{\sc{i}} recombination coefficients by
Benjamin et al. (1999) with those by Bauman, Ferland, \& MacAdam
(2002) indicate differences up to a few percent for some He~{\sc{i}}
line intensities. Similarly a comparison of the collisional
contribution to the helium lines by Benjamin et al. (1999) with those
by Porter (2002) indicate differences up to a few percent for some
He~{\sc{i}} line intensities.

The reddening correction can be systematically overestimated if the
collisional excitation of the H~{\sc{i}} lines is not taken into account,
introducing an additional correction to the derived He abundances
(Luridiana et al. 2003). 

Cota \& Ferland (1988) have argued that dusty H~{\sc{ii}} regions
might show deviations from Case B in the H~{\sc{i}} lines; this
effect, if present, will lower the derived He/H value. 

Finally the errors in the derivation of the line intensities have to
be considered: calibration of the standard stars, photon statistics,
atmospheric extinction, properties of the detector, etc.

\begin{table}
\begin{center}
\caption{Error Budget in the $Y_p$ Determination, given in 1/10000 of 
the mass fraction.}
\begin{tabular}{lcrccc@{$\pm$}rc}
\\
\hline
\hline
\rule[-3mm]{0mm}{8mm}
Problem & \multicolumn{3}{c}{Uncorrected} & \multicolumn{4}{c}{Corrected}\\ [0.2ex]
\hline
\\
Underlying Absorption in H~{\sc{i}} Lines       && $  -50$ &&&&  5 &\\
Underlying Absorption in He~{\sc{i}} Lines      && $  +70$ &&&&  7 &\\
He~{\sc{i}} and H~{\sc{i}} Line Intensities     &&    ...  &&&&  2 &\\
Ionization Structure                            && $\pm12$ &&&&  5 &\\
Temperature Structure                           && $  -60$ &&&& 15 &\\
Density Structure                               && $\pm45$ &&&& 10 &\\
Collisional Excitation of the He~{\sc{i}} Lines && $  -90$ &&&&  7 &\\
Collisional Excitation of the H~{\sc{i}} Lines  && $  +50$ &&&& 20 &\\
Optical Depth of the He~{\sc{i}} Triplet Lines  && $\pm10$ &&&&  3 &\\
He~{\sc{i}} and H~{\sc{i}} Atomic Parameters    && $\pm30$ &&&& 15 &\\
$\Delta Y/\Delta O$                             &&    ...  &&&& 10 &\\
\\
\hline
\hline
\end{tabular}
\end{center}
\end{table}

\section{Discussion and Conclusions}

In Table 2 we present our error estimates in the determination of
$Y_p$ from a typical set of well observed H~{\sc{ii}} regions.  In the
second column we present the size and sign of the bias on the $Y_p$
determination when the problem is ignored, for some problems we only
present the typical size because the sign depends on the specific
sample.  The third column includes the statistical errors; if the
problem is properly taken into account there should be no bias
present.  It is clear that the errors for a given determination will
depend on the included H~{\sc{ii}} regions. For example if the objects
are of relatively low $T_e$ the collisional excitation contribution to
the He~{\sc{i}} and H~{\sc{i}} lines will be small, and consequently
the expected errors due to collisional excitations; on the other hand
the error due to the adopted $\Delta Y$/$\Delta O$ will be relatively
large.

\begin{table}
\begin{center}
\caption{Primordial Helium Abundance Values}
\begin{tabular}{ll@{\,$\pm$\,}l@{\,$\pm$\,}l}
\\
\hline
\hline
\rule[-3mm]{0mm}{8mm}
Source& \multicolumn{3}{c}{$Y_p$} \\ [0.2ex]
\hline
\\
Izotov et al. (1999), this work    & 0.2452 & 0.0015 & 0.0070  \\
Peimbert et al. (2002), this work  & 0.2374 & 0.0035 & 0.0010?  \\
Prediction (2006), this work       & 0.2??? & 0.0020 & 0.0005  \\
\\
\hline
\hline
\end{tabular}
\end{center}
\end{table}

To derive a very accurate $Y_p$ value it is necessary to minimize the
sources of error presented in Table 2. The best objects to determine
$Y_p$ should have the following characteristics: a) a high H$\beta$ 
equivalent width in emission, b) a high degree of ionization, c) a low density
to have a low contribution due to collisional effects, but high enough to be
relatively bright, and d)
a moderately low O/H value to have a small $\Delta Y$ correction.
The metal poorest H~{\sc{ii}} regions
might not be the best candidates to derive an accurate $Y_p$ value
because due to their high $T_e$ values the corrections
due to collisional effects are very large.

In Table 3 we present three values of $Y_p$ together with our estimates of
the statistical and systematic errors. The first one is from the work by
Izotov et al. (1999), the second one from Peimbert et al. (2002) and Peimbert
(2003), and the third one is our prediction for the near future.

\acknowledgements{We are grateful to Evan Skillman and Gary Steigman
for several fruitful discussions, and to our colleagues
from Granada for such enjoyable meeting.}

\section*{\bf Discussion}

\noindent
{\it Evan Skillman:}

Keith Olive \& I have analyses which use the helium line strengths to
solve simultaneously for helium abundances, $\tau_{3889}$, density,
underlying absorption, and temperature; we reproduce your result that
the appropriate electron temperature is lower than the [O~{\sc{iii}}]
temperature, resulting in lower He abundances.


\begin{references}
 
\reference Armour, M. H., Ballantyne, D. R., Ferland, G. F.,
Karr., J, \& Martin, P. G. 1999, 
\pasp, 111, 1251
                    
\reference Ballantyne, D. R., Ferland, G. J., \& Martin, P. G. 2000,
\apj, 536, 773

\reference Bauman, R. P., Ferland, G. J., \& MacAdam, K. B. 2002,
in preparation

\reference Benjamin, R. A., Skillman, E. D., \& Smits, D. P. 1999,
\apj, 514, 307
 
\reference Benjamin, R. A., Skillman, E. D., \& Smits, D. P. 2002,
\apj, 569, 288

\reference Carigi, L. 2000, 
RevMexAA, 36, 171

\reference Carigi, L., Col\'{\i}n, P., \& Peimbert, M. 1999, 
\apj, 514, 787

\reference Carigi, L., Col\'{\i}n, P., Peimbert, M., \& Sarmiento, A. 1995, 
\apj, 445, 98

\reference Chiappini, C., Matteucci, F., \& Gratton, R. 1997, 
\apj, 477, 765

\reference Cota, S. A., \& Ferland, G. J. 1988,
\apj, 514, 787326, 889

\reference Davidson, K., \& Kinman, T. D. 1985,
\apjs, 58, 321

\reference Esteban, C., Peimbert, M., Torres-Peimbert, S., 
\& Escalante, V. 1998, 
\mnras, 295, 401

\reference Esteban, C., Peimbert, M., Torres-Peimbert, S., 
 \& Garc\'{\i}a-Rojas J. 1999,
RevMexAA, 35, 65

\reference Esteban, C., Peimbert, M., Torres-Peimbert, S., 
\& Rodr\'{\i}guez, M. 2002,
\apj, in press (astro-ph/0208313)

\reference Ferland, G. J., Korista, K. T., Verner, D. A., Ferguson, J. W., 
Kingdon, J. B., \& Verner, E. M. 1998,
\pasp, 110, 761

\reference Gonz\'alez-Delgado, R. M., Leitherer, C., \& Heckman, T. M. 1999,
\apjs, 125, 489

\reference Gruenwald, R., Steigman, G., \& Viegas, S. M. 2002, 
\apj, 567, 931

\reference Gruenwald, R., \& Viegas, S.M. 1992, 
\apjs, 78, 153

\reference Izotov, Y. I., Chaffee, F. H., Foltz, C. B., Green, R. F., 
Guseva, N. G., \&  Thuan, T. X. 1999,
\apj, 527, 757

\reference Izotov, Y. I., \& Thuan, T. X. 1998,
\apj, 500, 188

\reference Kingdon, J., \& Ferland, G. 1995,
\apj, 442, 714

\reference Keenan, F. P., Aller, L. H., Ryans, R. S. I., \& Hyung, S. 2001,
P. Natl. Acad. Sci. USA, 98, 9476

\reference Luridiana, V. 2002,
proceedings of the XXXVIIth Moriond Astrophysics Meeting "The Cosmological 
Model", in press, (astro-ph/0209177)

\reference Luridiana, V., Peimbert, M., \& Leitherer, C. 1999,
\apj, 527, 110

\reference Luridiana, V., Peimbert, A., \& Peimbert, M. 2003,
in preparation

\reference Pagel, B. E. J., Simonson, E. A., Terlevich, R. J., \& Edmunds, 
M. G. 1992, 
\mnras, 255, 325
 
\reference Peimbert, A. 2003,
\apj, in press, (astro-ph/0208502)

\reference Peimbert, A., Peimbert, M., \& Luridiana, V. 2002,
\apj, 565, 688

\reference Peimbert, M., \& Peimbert, A. 2000,
in IAU Symp. 198, The Light Elements and their Evolution, ed. L. da
Silva, M. Spite, \& J. R. de Medeiros (San Francisco: ASP), 194

\reference Peimbert, M., Peimbert, A., \& Ruiz, M. T. 2000,
\apj, 541, 688

\reference Peimbert, M., Storey, P. J., \& Torres-Peimbert, S. 1993, 
\apj, 414, 626

\reference Peimbert, M., Torres-Peimbert, S., \&  Ruiz, M. T. 1992,
RevMexAA, 24, 155

\reference Pe\~na, M. 1986, 
PASP, 98, 1061

\reference P\'erez, E. 1997, 
\mnras, 290, 465

\reference P\'erez, E., Gonz\'alez-Delgado, R. M., \& V\'{\i}lchez, J. 2001,
Ap. \& Sp. Sci., 277, 83

\reference Porter, R. L. 2002, in preparation

\reference Rela\~no, M., Peimbert, M., \& Beckman, J. 2002, 
\apj, 564, 704 

\reference Sauer, D., \& Jedamzik, K. 2002,
\aap, 381, 361

\reference Shields, G. A. 1974,
\apj, 191, 309

\reference Skillman, E. D., \& Kennicutt, E. D. 1993, ApJ, 411, 655

\reference Steigman, G. 2002,
XIII Canary Islands Winter School of Astrophysics, ``Cosmochemistry: The 
Melting Pot of Elements''(astro-ph/0208186)

\reference Stasi\'nska, G. 1983,
in ESO Workshop on Primordial Helium, 
ed. P. A. Shaver, D. Kunth, \& K. Kjar (Garching:ESO), p. 255

\reference Stasi\'nska, G. 1990,
A\&AS, 83, 501

\reference Stasi\'nska, G., \& Izotov, Y. I. 2001,
\aap, 378, 817

\reference Stasi\'nska, G., \& Schaerer, D. 1999,
\aap, 351, 72

\reference Terlevich, E., Terlevich, R. J., Skillman, E. D., Stepanian, E., 
\& Lipovetskii, V. 1991a, in Elements and the Cosmos, ed. M. G. Edmunds \& 
R. J. Terlevich (Cambridge: Cambridge Univ. Press), 21

\reference Terlevich, R. J., Melnick, J., Masegosa, J., \& Moles, M. 1991b, 
A\&AS, 91, 285

\reference Tsamis, Y. G., Barlow, M. J., Liu, X.-W., Danziger, I. J.,
\& Storey, P. J. 2003,
\mnras, in press (astro-ph/0209534) 

\reference Viegas, S. M., \& Gruenwald, R. 2000,
in IAU Symp. 198, The Light Elements and Their Evolution,
ed. L. da Silva, M. Spite, \& J. R. de Medeiros (San Francisco:ASP), 188
 
\reference Viegas, S. M., Gruenwald, R., \& Steigman, G. 2000, 
\apj, 531, 813

\reference V\'{\i}lchez, J. M., \& Pagel, B. E. J. 1988, 
\mnras, 231, 257

\reference Zurita, A., Rozas, M., \& Beckman, J. E. 2000, 
\aap, 363, 9


\end{references}
\end{document}